\documentclass[12pt]{article}
\pdfoutput=1
\usepackage{pub}

\usepackage{amsmath,bbm,array,amsfonts,graphicx,wrapfig,lscape,float,slashbox,multirow,longtable,rotating,subfigure}

\newcommand{\be}{\begin{equation}}
\newcommand{\ee}{\end{equation}}
\newcommand{\beq}{\begin{equation}}
\newcommand{\beql}[1]{\begin{equation}\label{#1}}
\newcommand{\eeq}{\end{equation}}
\newcommand{\ba}{\begin{array}}
\newcommand{\ea}{\end{array}}
\newcommand{\bea}{\begin{eqnarray}}
\newcommand{\beal}[1]{\begin{eqnarray}\label{#1}}
\newcommand{\eea}{\end{eqnarray}}
\newcommand{\ben}{\begin{enumerate}}
\newcommand{\een}{\end{enumerate}}
\newcommand{\bean}{\begin{eqnarray*}}
\newcommand{\eean}{\end{eqnarray*}}
\newcommand{\eref}[1]{(\ref{#1})}
\newcommand{\sref}[1]{\S\ref{#1}}

\newcommand{\fref}[1]{Figure \ref{#1}}
\newcommand{\btab}[1]{\begin{tabular}{#1}}
\newcommand{\etab}{\end{tabular}}

\newcommand{\comment}[1]{}

\newcommand{\qed}{\nobreak \ifvmode \relax \else
      \ifdim\lastskip<1.5em \hskip-\lastskip
      \hskip1.5em plus0em minus0.5em \fi \nobreak
      \vrule height0.75em width0.5em depth0.25em\fi}

\newcolumntype{C}[1]{>{\centering\arraybackslash}m{#1}}

\newcommand{\bb}{\mathsf{b}}
\newcommand{\ww}{\mathsf{w}}

\title{Cluster Transformations from Bipartite Field Theories}

\author{Sebasti\'an Franco}

\affiliation{
Institute for Particle Physics Phenomenology, Department of Physics\\
Durham University, Durham DH1 3LE, United Kingdom
}

\emailAdd{sebastian.franco@durham.ac.uk}

\abstract{Bipartite field theories (BFTs) are a new class of 4d $\mathcal{N}=1$ quantum field theories defined by bipartite graphs on bordered Riemann surfaces. In this paper we derive, purely in terms of the gauge theory, the cluster transformations of face weights under square moves in the graph. In this context, we obtain them by connecting regular parametrizations of the master space of the associated BFTs. For BFTs on a disk, these transformations follow from the properties of coordinates in the Grassmannian. This represents a new addition to the list of combinatorial objects for the Grassmannian, such as matching and matroid polytopes, that have been shown to emerge from BFT dynamics.
}

\preprint{
\begin{flushright}IPPP/12/98\end{flushright} \vspace{-0.9cm}
\begin{flushright}DCPT/12/196\end{flushright}
}

\begin{document}

\maketitle

\section{Introduction}

Bipartite field theories (BFTs) are a class of 4d $\mathcal{N}=1$ gauge theories defined by bipartite graphs on Riemann surfaces, which might have borders \cite{Franco:2012mm,Xie:2012mr}. This class of theories contains and generalizes brane tilings on a 2-torus \cite{Hanany:2005ve,Franco:2005rj}.

BFTs are certainly interesting in their own right. In addition, they provide useful intuition and insights, based on standard quantum field theory, into other problems related to bipartite graphs. Examples of such systems with a one-to-one correspondence with BFTs include: D-branes over toric Calabi-Yau (CY) 3-folds \cite{Hanany:2005ve,Franco:2005rj,Franco:2005sm,Butti:2005sw,Feng:2005gw}, cluster integrable systems \cite{GK,Franco:2011sz,Eager:2011dp,Franco:2012hv} and scattering amplitudes in $\mathcal{N}=4$ SYM \cite{Nima}. Cross-fertilization works in both directions, and these systems can be also exploited for gaining a deeper understanding of BFTs.

One reason connecting scattering amplitudes in planar $\mathcal{N}=4$ SYM and bipartite graphs is that both sets of objects are related to the cell decomposition of the totally nonnegative Grassmannian \cite{Nima,ArkaniHamed:2009dn,Postnikov_plabic,Postnikov_toric}. It then becomes natural to try to uncover objects and properties that are important for the classification of cells in the Grassmannian in terms of BFTs. Considerable progress has been made on this front since the recent inception of BFTs. Among other things, cells in the Grassmannian have been linked to low energy equivalence classes of Seiberg dual theories \cite{Franco:2012mm,Xie:2012mr}, the boundary operator on them has been interpreted as higgsing \cite{Franco:2012mm}, the matching and matroid polytopes have been identified with the toric diagrams of the master space and the mesonic moduli space \cite{Franco:2012mm,Franco:2012wv}, and cluster transformations of face weights \cite{Postnikov_plabic} have been derived in terms of vevs of line operators in the dimensional reduction of BFTs to 3d \cite{Heckman:2012jh}. The aim of this paper is to extend this list by obtaining the cluster transformations of face weights directly at the level of the BFT. We will show that cluster transformations connect regular parametrizations of the master spaces of theories related by square moves. In the context of scattering amplitudes, cluster transformations give rise to useful changes of variables in on-shell forms. Furthermore, in combination with graph reduction, they can be used to simplify the determination of leading singularities \cite{Nima}.

In contrast with our field theoretic approach, the discussion of cluster transformations for bipartite graphs is often phrased in terms of rather abstract concepts such as the partition function of a dimer model \cite{Propp} or the invariance of the boundary measurement \cite{Postnikov_plabic}. Given the recent irruption of these ideas in the physics literature, our primary aim is to provide an alternative understanding of them in terms of objects that are closer to physicists. We certainly do not claim a first time derivation of cluster transformations, which were originally found in \cite{FZ4}. We hope our presentation clarifies some subtle points to non-experts and provides another handle on cluster transformations to the physically inclined reader.

Further developments in the study of BFTs include the connection to toric CY manifolds, a detailed study of non-planar BFTs, the relation to BPS quiver of 4d $\mathcal{N}=2$ gauge theories and a string theory realization of some BFTs in terms of D5 and NS5-branes \cite{Franco:2012mm,Franco:2012wv, Heckman:2012jh}.

This paper is organized as follows. \sref{section_BFTs_nutshell} briefly reviews BFTs and the computation of their master and moduli spaces. \sref{section_square_moves} discusses the BFT interpretation of square moves. Cluster transformations for face weights are derived from regular parametrizations of BFT master spaces in \sref{section_electric_magnetic}. We conclude in \sref{section_conclusions}. Appendix \ref{appendix_legs} discusses the BFT treatment of external legs of bipartite graphs.

\bigskip

\section{BFTs in a Nutshell}

\label{section_BFTs_nutshell}

Bipartite Field Theories are a general class of 4d, $\mathcal{N} = 1$ quiver gauge theories whose Lagrangians are defined by bipartite graphs on Riemann surfaces, with or without boundaries \cite{Franco:2012mm,Xie:2012mr}. Table \ref{tdic} summarizes the dictionary connecting bipartite graphs on Riemann surfaces and BFTs.

\begin{table}[htt!!]
\begin{center}
\begin{tabular}{|l|l|}
\hline
{\bf Graph} & {\bf BFT} \\ \hline \hline
Internal face ($2n$-sided) & Gauge group with $n$ flavors \\ \hline
External face & Global symmetry group \\ \hline \hline
Edge between two faces $i$ and $j$ & Chiral multiplet in the bifundamental \\ & representation of the groups $i$ and $j$. The \\ & orientation of the corresponding arrow is such \\ & that it goes clockwise around white nodes and \\ & counterclockwise around black nodes. \\ \hline \hline
$k$-valent node & Monomial in the superpotential involving $k$ \\ & multiplets. The signs of the terms are \\ & (+/-) for (white/black) nodes. External nodes \\ & do not correspond to superpotential terms. \\ \hline
\end{tabular}
\caption{The dictionary connecting bipartite graphs on Riemann surfaces and BFTs. \label{tdic}
}
\end{center}
\end{table}

\begin{figure}[h]
\begin{center}
\includegraphics[width=12cm]{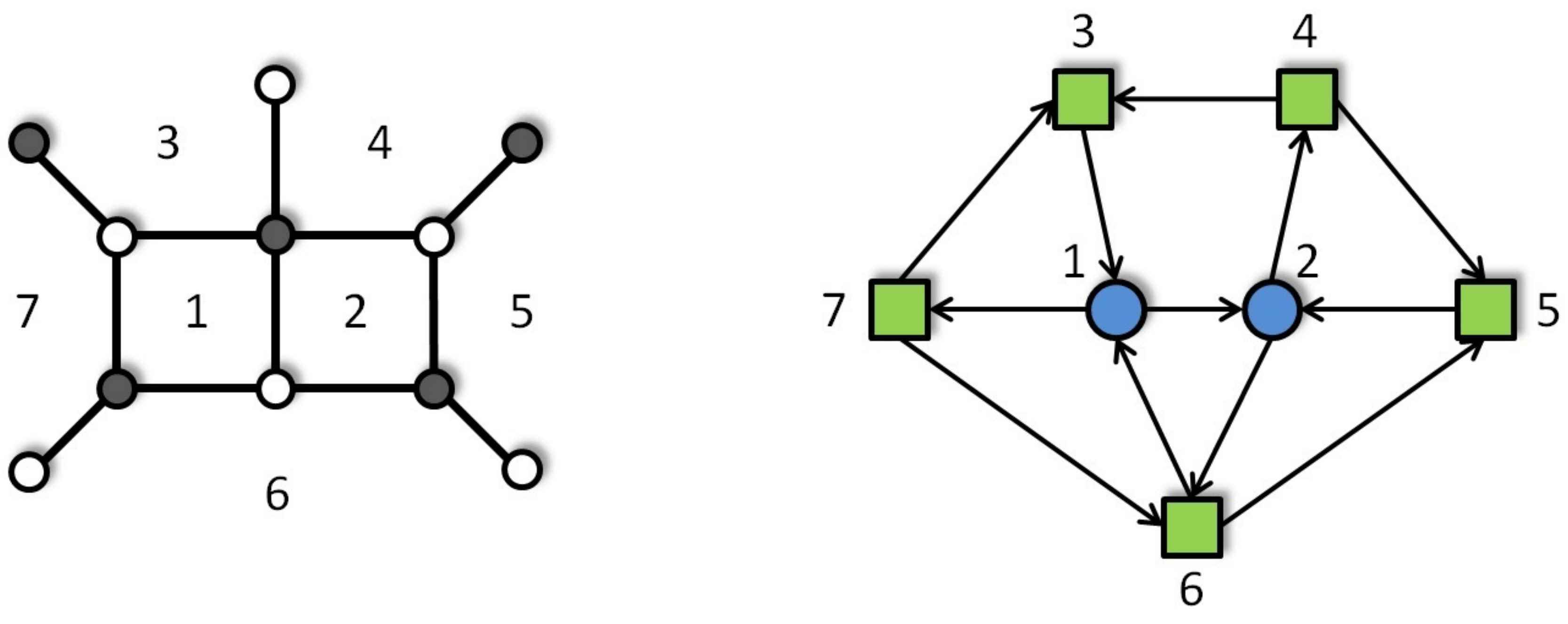} \vspace{-.3cm}
\caption{\textit{Bipartite graph and its dual BFT.} This example corresponds to the top dimensional cell of $G(2,5)$.}
\label{tiling_quiver_G25}
\end{center}
\end{figure}

We refer the reader to \cite{Franco:2012mm} for a detailed explanation of the correspondence. The quiver for a BFT lives on the Riemann surface and is obtained by dualizing the defining bipartite graph. Plaquettes in this quiver encode superpotential terms. The gauge and global symmetry groups are $U(N_i)$, where the index $i$ runs over all faces of the graph. The ranks $N_i$ are constrained by anomaly cancellation. For gauge groups, the central $U(1) \subset U(N_i)$ flow to zero coupling at low energies and become global symmetries. \fref{tiling_quiver_G25} shows an example of a bipartite graph and its corresponding BFT.

We will regard the chiral fields associated to edges connected to external nodes of the graph as non-dynamical. This implies that we neither impose vanishing of their F-terms nor integrate them out when they participate in mass terms. In the latter case, we just consistently keep them in our analysis. There are various reasons motivating this choice, which have been already discussed in \cite{Franco:2012mm,Franco:2012wv}. First, naive promotion of these fields, which participate in a single superpotential term, to dynamical objects severely constrains the moduli space of BFTs. In addition, in certain D-brane realizations of some of these theories, external faces correspond to higher dimensional branes such as D7-branes. The fields associated to external legs live at the intersections between these flavor branes, which are infinite along the internal dimensions. As a result, they are non-dynamical from a 4d viewpoint. Finally, and perhaps more importantly, this treatment beautifully leads to the emergence of various objects in the combinatorics of the Grassmannian from BFT dynamics. Examples include the matching and matroid polytopes \cite{Postnikov_plabic,Postnikov_toric}, which are linked to the master and moduli spaces of the BFT \cite{Franco:2012mm,Franco:2012wv}.

Other consistent treatments of external legs are possible, such as the one considered in \cite{Xie:2012mr}. In Appendix \ref{appendix_legs} we collect some general remarks on the connection between different approaches.

In \cite{Franco:2012wv}, an alternative way of gauging anomaly free symmetries in these models was identified, giving rise to a new class of BFTs. The discussion in the following sections extends to those theories with small changes. For simplicity, this extended gauging will not be considered in this paper.

\bigskip

\subsection{Moduli Spaces of BFTs}

\label{section_moduli_space_BFT}

The {\bf mesonic moduli space} of a general $\mathcal{N}=1$ gauge theory, or {\bf moduli space} for short, is the space of solutions to vanishing F and D-terms. The {\bf master space} is given by the solutions to only vanishing F-terms \cite{Forcella:2008bb}, and can be regarded as an intermediate step in the determination of the moduli space. 

In this article we focus on the moduli space of Abelian BFTs. When doing so, we consider the classical theory, in which the gauge couplings are constant and non-zero instead of vanishing at low energies. The main reason for doing so is that the Abelian case is directly relevant for scattering amplitudes. An Abelian BFT can be mapped to a $U(1)$ gauge theory living on the graph \cite{Kenyon:2003uj,Franco:2006gc}, which also provides an alternative formulation of the scattering problem \cite{Nima}.\footnote{Despite the connection between the two theories, the $U(1)$ gauge theory on the graph should not be confused with the corresponding Abelian BFT.} In addition, the Abelian case is sufficient for capturing the equivalence graph equivalences that correspond to Seiberg duality in the non-Abelian case. It is reasonable to expect that the moduli space of a non-Abelian theory with all ranks equal to $N$ is related to the symmetric product of $N$ copies of the Abelian moduli space. This is indeed the case for BFTs with a D-brane interpretation. A more general investigation of generic BFTs is desirable in order to determine whether such simple connection holds more generally, although it is beyond the scope of this paper. The only potential cause of differences with the well-understood D-brane case is given by global properties of the graph. In any case, since Seiberg duality acts as a local modification of the graph, our discussion can be promoted to non-Abelian theories without changes.

When parametrizing the moduli space in terms of the scalar components of the chiral fields in the theory, expressions involving negative powers are generically encountered. We refer to such parametrizations as {\bf irregular}, since they contain poles at the origin of some directions in field space. Explicit examples illustrating this situation are given in \sref{section_electric_magnetic}. It is thus natural to ask whether it is possible to find a {\bf regular parametrization} of the master space. By this we mean a parametrization in terms of a new set of fields $p_\mu$ such that, on the master space, all $X_i$ can be expressed as

\beq
X_i(p_\mu)=\prod_{\mu=1}^c p_\mu ^{P_{i\mu}},
\label{X_of_p}
\eeq
where $P$ is a matrix with integer entries greater or equal to zero. The same information can be alternatively encoded in terms of a {\bf characteristic polynomial} $\mathcal{P}$, in which each term $\mathcal{P}_\mu$ is in one-to-one correspondence with a field $p_\mu$:
\beq
\mathcal{P} = \sum_{\mu=1}^c \mathcal{P}_\mu, \ \ \ \ \ \ \ \ \mathcal{P}_\mu (X_i)=\prod_i X_i^{P_{i\mu}}.
\label{p_of_X}
\eeq
While there is a one-to one correspondence between $p_\mu$ and $\mathcal{P}_\mu$, there is a subtle difference. The $X_i$ fields should be understood as functions of the $p_\mu$, but $\mathcal{P}_\mu(X_i)$ is not the result of inverting this map, which in fact is not invertible. In \sref{section_connecting_parametrizations} we provide further details on the interpretation of \eref{p_of_X}.

As a result of the restricted structure of BFT theories, whose Lagrangian is dictated by a bipartite graph, it is always possible to find a regular parametrization of their master space. Moreover, the fields $p_\mu$ have a graphical representation and are identified with {\bf almost perfect matchings} of the bipartite graph. An almost perfect matching, or {\bf perfect matching} for brevity, $p$ is a subset of the edges in the graph such that: 1) every internal node is the endpoint of exactly one edge in $p$ and 2) every external node belongs to either one or zero edges in $p$. In fact, $P_{i\mu}$ can be read directly from the graph and is equal to $1$ if the edge associated to the chiral field $X_i$ is contained in $p_\mu$ and zero otherwise, i.e.

\beq
P_{i\mu}=\left\{ \begin{array}{ccccc} 1 & \rm{ if } & X_i  & \in & p_\mu \\
0 & \rm{ if } & X_i  & \notin & p_\mu
\end{array}\right.
\label{Xi_to_pmu}
\eeq

\begin{figure}[h]
\begin{center}
\includegraphics[width=8.5cm]{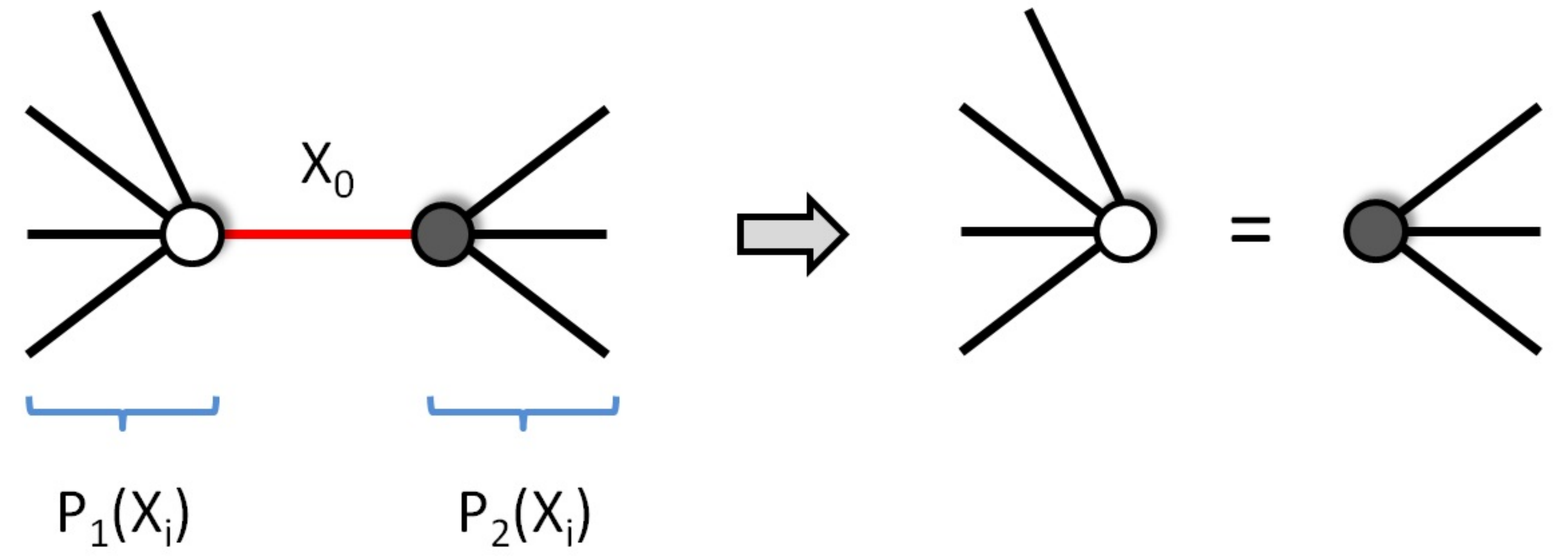}
\caption{\textit{Graphic representation of the F-term equations in a BFT.}}
\label{F_terms_graphical}
\end{center}
\end{figure}

The map between chiral fields and perfect matchings determined by \eref{X_of_p} and \eref{Xi_to_pmu} implies that F-term equations are automatically satisfied \cite{Franco:2005rj,Franco:2012mm}. This can be understood as follows. Every bifundamental field $X_0$ associated to an internal edge appears in exactly two terms in the superpotential, i.e. 
\beq
W=X_0 P_1(X_i)- X_0 P_2(X_i) +\ldots,
\eeq
where $P_1(X_i)$ and  $P_2(X_i)$ are products of bifundamentals fields. The F-term equation for $X_0$ is given by

\beq
\partial_{X_0}W =0 \ \ \ \iff \ \ \ P_1(X_i)=P_2(X_i).
\label{F_internal}
\eeq
\fref{F_terms_graphical} shows the graphical representation of this equation. Removing the edge associated to $X_0$, the product of edges connected to the node on the left needs to be equal to the product of edges connected to the node on the right. Using \eref{X_of_p}, the F-term equation takes the general form

\beq
\prod_{i \in P_1} \prod_{\mu=1}^c p_\mu^{P_{i\mu}}=\prod_{i \in P_2} \prod_{\mu=1}^c p_\mu^{P_{i\mu}}.
\label{F_terms_pms}
\eeq
Since the two nodes under consideration are separated by a single edge, all perfect matchings that appear on the L.H.S. of \eref{F_terms_pms} also appear on its R.H.S.. We thus conclude all F-term equations are satisfied.

The next step for determining the moduli space is to impose vanishing of D-terms. As already said, we focus on Abelian BFTs. The transformation properties under gauge symmetries of the $X_i$ fields are implemented by assigning non-trivial charges to the $p_\mu$ fields. Following \eref{X_of_p}, for each factor $U(1)^{(\alpha)}$ of the gauge group, the corresponding charges are given by 
\beq
Q^{(\alpha)}(X_i)=\sum_{\mu=1}^c  P_{i\mu} \, Q^{(\alpha)}(p_\mu).
\eeq
These equations are used for determining an assignation of charges $Q^{(\alpha)}(p_\mu)$ that is consistent with the values of $Q^{(\alpha)}(X_i)$ for every $i$ and $\alpha$. Such charges are typically not unique, but the resulting moduli space is independent of how they are chosen. Gauge invariant operators parametrizing the moduli space correspond to products of perfect matchings that are neutral under all gauge symmetries. Extending what they do for the master space, perfect matching of course give rise to a regular parametrization of the moduli space.

The master and moduli spaces of Abelian BFTs are toric CY manifolds \cite{Franco:2012mm,Franco:2012wv}. The construction we have outlined beautifully fits into the gauged linear sigma model (GLSM) realization of these geometries. In this note, however, we will mainly focus on the field theory side of the story. It is important to emphasize that while the $p_\mu$ fields have a natural interpretation as perfect matchings of a graph, the existence and determination of a regular parametrization of the moduli space of a BFT can be entirely phrased in gauge theoretic terms.

The discussion in this section is sufficient for the purposes of this article. We refer the reader to \cite{Franco:2012mm,Franco:2012wv} for further details on the calculation of master and moduli spaces for BFTs.

\bigskip

\section{Square Moves}

\label{section_square_moves}

\begin{figure}[h]
\begin{center}
\includegraphics[width=10cm]{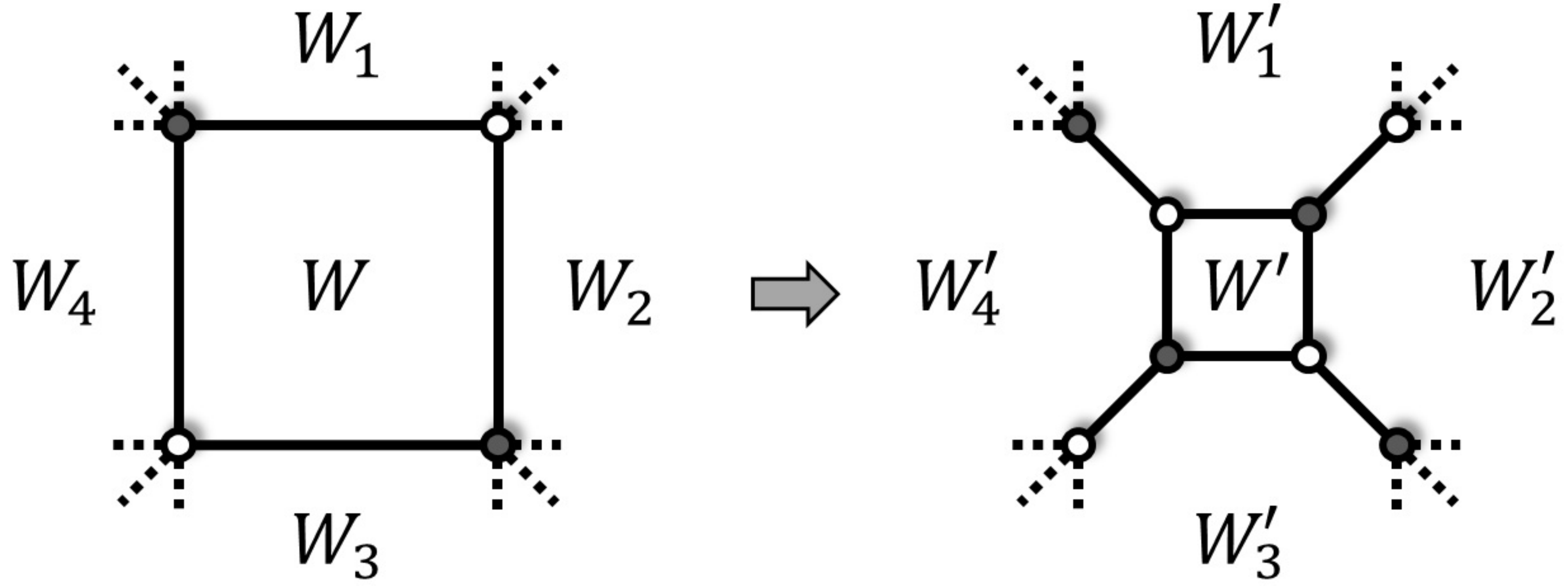}
\caption{\textit{Square move in a bipartite graph.} We label the faces affected by the move, whose transformation properties will be later determined.}
\label{square_move_faces}
\end{center}
\end{figure}

In this article we will be interested in the behavior of BFTs under the {\bf square move} transformation, also referred to as {\bf urban renewal}, shown in \fref{square_move_faces}. This operation can be applied to any internal square face of the graph. For non-Abelian BFTs, it corresponds to Seiberg duality \cite{Seiberg:1994pq} of the corresponding gauge group \cite{Franco:2005rj,Franco:2012mm,Xie:2012mr}. The square move replaces electric quarks with magnetic quarks, introduces Seiberg mesons and generates cubic superpotential couplings between mesons and magnetic quarks. When the ranks of all faces are equal, gauge groups associated to square faces have $N_f=2N_c$ flavors.

The moduli space is invariant under square moves for both non-Abelian BFTs, as a consequence of Seiberg duality, and classical Abelian BFTs.

\bigskip

\section{Regular Parametrizations of the Master Space and Cluster Transformations}

\label{section_electric_magnetic}

As explained above, every BFT admits a regular parametrization of its master space. Furthermore, the fields in this parametrization are in one-to-one correspondence with perfect matchings of the underlying bipartite graph. We now investigate the effect of square moves of the graph on regular parametrizations. As already mentioned, we will focus on Abelian BFTs. 

Motivated by the standard nomenclature, we refer to the theories before and after the square move as {\bf electric} and {\bf magnetic}, respectively. While it is certainly straightforward to determine the perfect matching parametrizations of the master spaces of both the electric and magnetic BFTs, it is natural to ask whether there is a more efficient procedure that directly produces the perfect matching parametrization for one of the theories starting from the one for its dual. In this section we explain how {\bf cluster transformations} of face weights \cite{Postnikov_plabic} provide such a map. Cluster algebras were introduced in \cite{MR1887642}.

Since a square move is a local transformation of the graph, it is sufficient to focus on its effect on the immediate neighborhood of the dualized gauge group. For this reason, below we identify the {\bf minimal theory} accepting square moves that can be considered in the BFT context and discuss the field theory computation of its moduli space in detail.

Let us emphasize that cluster transformations are well understood for quiver nodes with an arbitrary number of arrows \cite{FZ4,MR1887642}. Similarly, Seiberg duality is completely clear for general $N_f>N_c$ \cite{Seiberg:1994pq}. The reason for focusing on square moves is that such transformations keep us within the class of theories which are described by bipartite graphs, which are the objects we are interested in studying. 

\bigskip

\subsection{The Electric Theory}

\begin{figure}[h]
\begin{center}
\includegraphics[width=5.3cm]{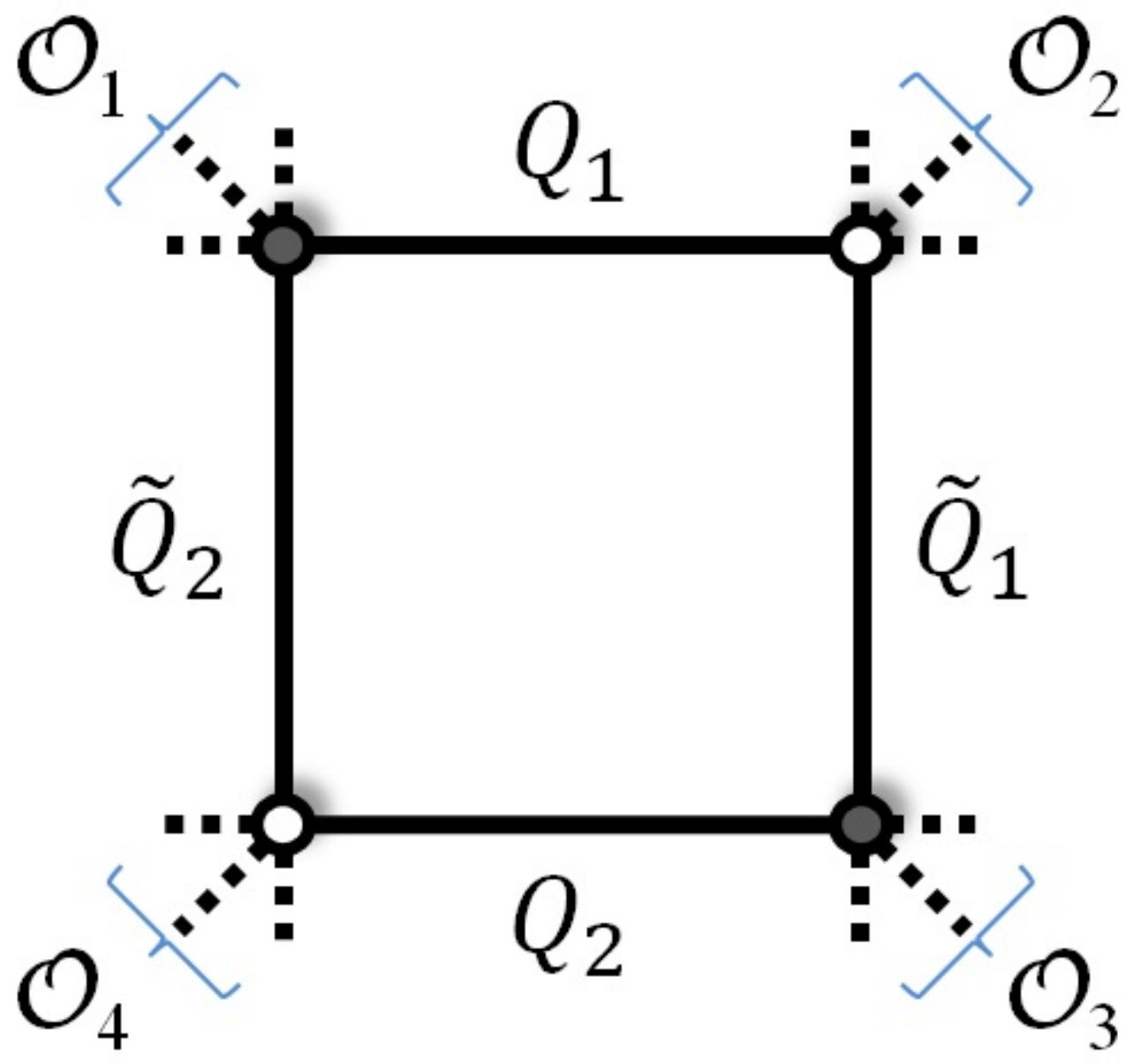}
\caption{\textit{The minimal electric configuration.}}
\label{tiling_minimal_electric}
\end{center}
\end{figure}

\fref{tiling_minimal_electric} shows the minimal theory, which can be regarded as a piece of a more complicated BFT. This diagram is interpreted as $N_f=2N_c$ SQCD augmented by gauge invariant, from the point of view of the dualized gauge group, operators $\mathcal{O}_\alpha$, $\alpha=1,\ldots,4$.\footnote{In the full theory, the faces adjacent to the dualized square might be closed and hence gauged.} The $2N_c$ flavors corresponds to chiral superfields $Q_i$ and $\tilde{Q}_i$, $i=1,2$ where, for simplicity, an additional flavor index coming from the fact that these fields are actually bifundamental has been omitted. In the Abelian case, this is a $U(1)$ gauge theory, with charge $1$ and $-1$ chiral fields $Q_i$ and $\tilde{Q}_i$, and the four $\mathcal{O}_\alpha$ operators. In addition, the theory has a the following superpotential

\beq
W_{el}=-\tilde{Q}_2 Q_1  \mathcal{O}_1 +\tilde{Q}_1 Q_1  \mathcal{O}_2 -\tilde{Q}_1 Q_2  \mathcal{O}_3 +\tilde{Q}_2 Q_2 \mathcal{O}_4.
\label{W_el}
\eeq

This theory can be regarded as a piece of a larger BFT. In the full theory, the $\mathcal{O}_\alpha$ operators can correspond to either single chiral fields or products of them. Notice that it is not possible to consistently study just a square graph, i.e. without coupling it to the $\mathcal{O}_\alpha$ operators. Naively, one would think that such a configuration corresponds to pure $N_f=2N_c$ SQCD. In fact, the 2-valent nodes at the corners of the square correspond to quadratic terms in the superpotential giving mass to all the flavors. Integrating them out would trigger a process in which the entire square disappears.

When computing the moduli space of the gauge theory, we will give the $\mathcal{O}_\alpha$ operators a special treatment: their F-terms equations are not imposed. This choice follows from considerations that are similar to the ones mentioned in \sref{section_BFTs_nutshell} for regarding the chiral fields associated to external legs of bipartite graphs as non-dynamical \cite{Franco:2012mm, Franco:2012wv}. Intuitively, not imposing their F-terms amounts to considering a patch in the moduli space of the full theory.

Let us now focus on the Abelian theory and compute its moduli space. The first step is solving F-term equations, whose space of solutions corresponds to the master space of the theory. Vanishing of the F-terms for $Q_1$, $Q_2$, $\tilde{Q}_1$, $\tilde{Q}_2$ becomes

\beq
\begin{array}{ccccccc}
\mathcal{O}_2 \, \tilde{Q}_1 & = & \mathcal{O}_1 \, \tilde{Q}_2 & \ \ \ \ \ \ \ \ & \mathcal{O}_2 \, Q_1 & = & \mathcal{O}_3 \, Q_2 \\
\mathcal{O}_3 \, \tilde{Q}_1 & = & \mathcal{O}_4 \, \tilde{Q}_2 &            & \mathcal{O}_1 \, Q_1 & = & \mathcal{O}_4 \, Q_2
\end{array}
\label{F_electric}
\eeq
These equations can be solved in terms of a subset of the $\{Q_i,\tilde{Q}_i,\mathcal{O}_\alpha\}$ variables. For example, we can have: $\mathcal{O}_2 =(\mathcal{O}_1 \tilde{Q}_2)/\tilde{Q}_1$, $\mathcal{O}_3=(\mathcal{O}_1 Q_1 \tilde{Q}_2)/(\tilde{Q}_1 Q_2)$, $\mathcal{O}_4=(O_1 Q_1)/Q_2$. This is a singular parametrization, since it involves negative powers of fields.

It is straightforward to find a regular parametrization of the space of solutions of \sref{F_electric} without making any reference to the graph underlying the gauge theory. Following the general discussion in \sref{section_moduli_space_BFT}, it is instead possible to take a shortcut and exploit the fact that the $p_\mu$ fields are associated to perfect matchings. The theory at hand is so simple that all perfect matchings can be determined by direct inspection of the graph. Alternatively, one can use the systematic Kasteleyn matrix methods introduced in \cite{Franco:2012mm}. The perfect matching matrix $P$ connecting $p_\mu$ fields to chiral fields in the BFT takes the form

{\small
\beq
P_e=\left(
\begin{array}{c|ccccccc}
& \ p_1 \ & \ p_2 \ & \ p_3 \ & \ p_4 \ & \ p_5 \ & \ p_6 \ & \ p_7 \ \\ \hline
\ \mathcal{O}_1 \ \ & 1 & 1 & 1 & 0 & 0 & 0 & 0 \\
\ \mathcal{O}_2 \ \ & 1 & 1 & 0 & 1 & 0 & 0 & 0 \\
\ \mathcal{O}_3 \ \ & 1 & 0 & 0 & 1 & 1 & 0 & 0 \\
\ \mathcal{O}_4 \ \ & 1 & 0 & 1 & 0 & 1 & 0 & 0 \\
\ Q_1 \ \ & 0 & 0 & 0 & 0 & 1 & 1 & 0 \\
\ \tilde{Q}_1 \ \ & 0 & 0 & 1 & 0 & 0 & 0 & 1 \\
\ Q_2 \ \ & 0 & 1 & 0 & 0 & 0 & 1 & 0 \\
\ \tilde{Q}_2 \ \ & 0 & 0 & 0 & 1 & 0 & 0 & 1 
\end{array}
\right).
\label{P_matrix_electric}
\eeq}
\smallskip

\noindent One should not be scared by this matrix, its sole purpose is to encode a regular parametrization of the master space. For illustration, let us explicitly see how the perfect matching parametrization automatically satisfies the first F-term equation in \eref{F_electric}, i.e. $F_{Q_1}=0$.

\beq
\left. \begin{array}{ccccccc}
\mathcal{O}_1 & = & p_1 p_2 p_3 & , \ \ & \tilde{Q}_1 & = & p_3 p_7 \\
\mathcal{O}_2 & = & p_1 p_2 p_4 & , \ \ & \tilde{Q}_2 & = & p_4 p_7 
\end{array} \right\} \ \ F_{Q_1}: \ \ (p_1 p_2 p_4)(p_3 p_7) = (p_1 p_2 p_3)(p_4 p_7).
\eeq
All other F-term equations are satisfied in a similar way. As explained above, D-terms are satisfied by restricting to gauge invariant combinations of the $p_\mu$ fields.

The information in \eref{P_matrix_electric} can be alternatively encoded in terms of the characteristic polynomial

\beq
\mathcal{P}_e  = (Q_1 Q_2 + \tilde{Q}_1 \tilde{Q}_2) + \mathcal{O}_1 \mathcal{O}_2 Q_2 + \mathcal{O}_1 \mathcal{O}_4 \tilde{Q}_1 + \mathcal{O}_2 \mathcal{O}_3 \tilde{Q}_2 + \mathcal{O}_3 \mathcal{O}_4 Q_1+ \mathcal{O}_1 \mathcal{O}_2 \mathcal{O}_3 \mathcal{O}_4,
\label{polynomial_electric}
\eeq
in which each term corresponds to a $p_\mu$. We have organized terms according to the $\mathcal{O}_\alpha$ content instead of following the order of columns in the matrix $P_e$.

\bigskip

\subsection{The Magnetic Theory}

\begin{figure}[h]
\begin{center}
\includegraphics[width=5.3cm]{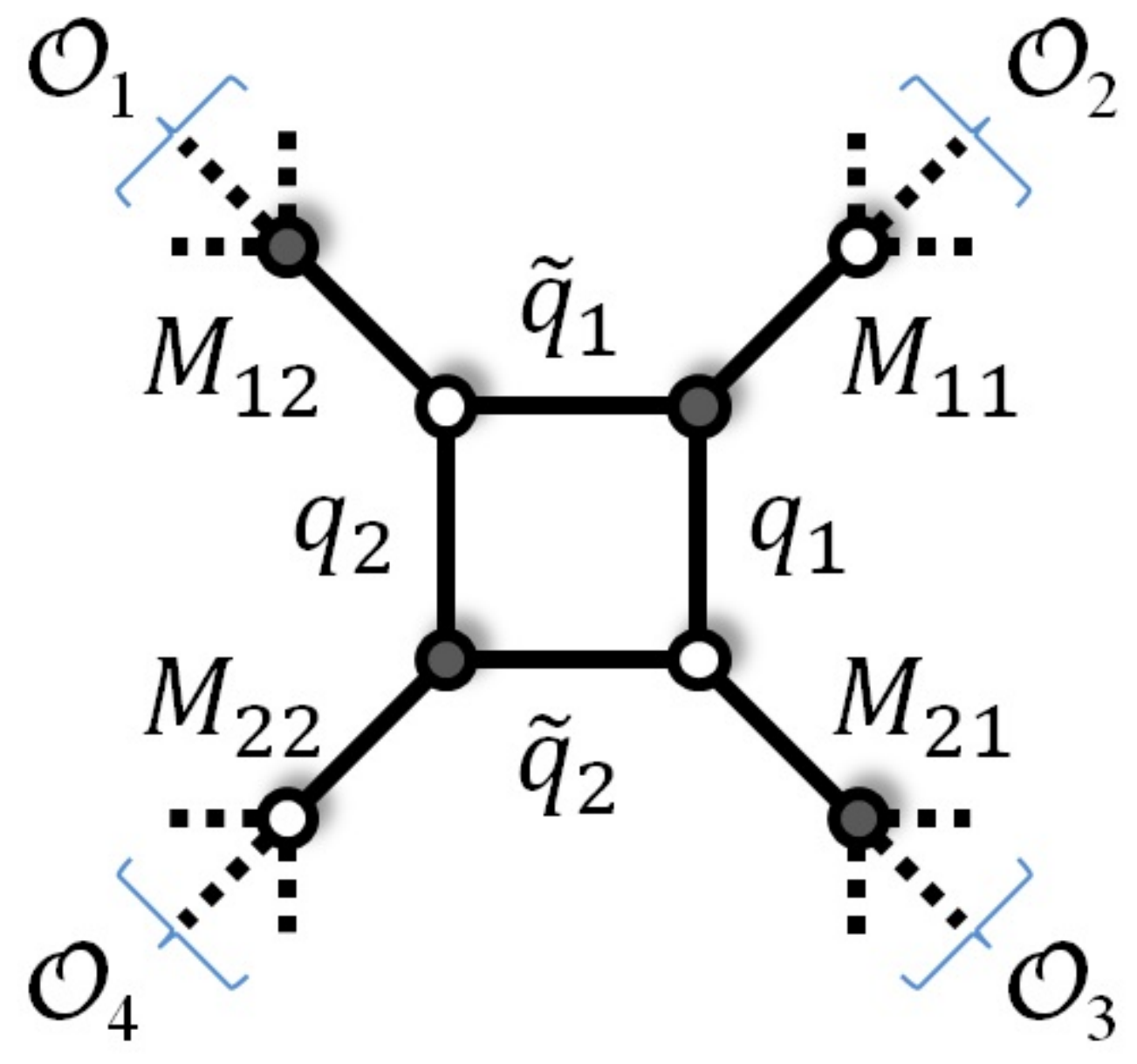}
\caption{\textit{The dual of the minimal electric configuration.}}
\label{tiling_minimal_magnetic}
\end{center}
\end{figure}

Let us analyze the theory shown in \fref{tiling_minimal_magnetic}, which is obtained by acting on the minimal theory with a square move. In the non-Abelian case, this model follows from a Seiberg duality on the gauge group associated to the square face. This theory has the matter content of usual magnetic SQCD, with the addition of the gauge invariants $\mathcal{O}_\alpha$ and superpotential

\beq
W_{mag}=[-M_{12}  \mathcal{O}_1 +M_{11}  \mathcal{O}_2 -M_{21}  \mathcal{O}_3 +M_{22} \mathcal{O}_4]+ \tilde{q}_1 q_2  M_{12} - \tilde{q}_1 q_1  M_{11}+ \tilde{q}_2 q_2  M_{22}- \tilde{q}_2 q_1  M_{21}, 
\label{W_mag}
\eeq
where the terms in square brackets come from the electric superpotential in \eref{W_el} and the rest are the usual cubic terms introduced by Seiberg duality.

Let us now focus on the Abelian version of this model. The F-term equations for the $q_i$, $\tilde{q}_i$ and $M_{ij}$ fields are

\beq
\begin{array}{ccccccc}
M_{21} \, \tilde{q}_2 & = & M_{11} \, \tilde{q}_1 & \ \ \ \ \ \ \ \ & \tilde{q}_1 \, q_2 & = & \mathcal{O}_1 \\
M_{12} \, q_2 & = & M_{11} \, q_1 & \ \ \ \ \ \ \ \ & q_1 \, \tilde{q}_1 & = & \mathcal{O}_2 \\
M_{12} \, \tilde{q}_1 & = & M_{22} \, \tilde{q}_2 & \ \ \ \ \ \ \ \ & q_1 \, \tilde{q}_2 & = & \mathcal{O}_3 \\
M_{21} \, q_1 & = & M_{22} \, q_2 & \ \ \ \ \ \ \ \ & q_2 \, \tilde{q}_2 & = & \mathcal{O}_4
\end{array}
\label{F_mag}
\eeq
A regular parametrization of the master space, the space of solutions to these equations, is summarized by the following matrix

{\small
\beq
P_m=\left(
\begin{array}{c|ccccccc}
& \ p'_1 \ & \ p'_2 \ & \ p'_3 \ & \ p'_4 \ & \ p'_5 \ & \ p'_6 \ & \ p'_7 \ \\ \hline
\ \mathcal{O}_1 \ \ & 0 & 0 & 1 & 0 & 1 & 1 & 1 \\
\ \mathcal{O}_2 \ \ & 0 & 0 & 0 & 1 & 1 & 1 & 1 \\
\ \mathcal{O}_3 \ \ & 0 & 1 & 0 & 1 & 0 & 1 & 1 \\
\ \mathcal{O}_4 \ \ & 0 & 1 & 1 & 0 & 0 & 1 & 1 \\
\ M_{11} \ \ & 1 & 1 & 1 & 0 & 0 & 0 & 0 \\
\ M_{12} \ \ &  1 & 1 & 0 & 1 & 0 & 0 & 0 \\
\ M_{21} \ \ & 1 & 0 & 1 & 0 & 1 & 0 & 0 \\
\ M_{22} \ \ & 1 & 0 & 0 & 1 & 1 & 0 & 0 \\
\ q_1 \ \ & 0 & 0 & 0 & 1 & 0 & 1 & 0 \\
\tilde{q}_1 \ \ & 0 & 0 & 0 & 0 & 1 & 0 & 1 \\
\ q_2 \ \ & 0 & 0 & 1 & 0 & 0 & 1 & 0 \\
\ \tilde{q}_2 \ \ & 0 & 1 & 0 & 0 & 0 & 0 & 1
\end{array}
\right).
\label{P_magnetic}
\eeq}

\smallskip

\noindent For example, for $F_{q_1}$ we have

\beq
\left. \begin{array}{ccccccc}
M_{21} & = & p'_1 p'_3 p'_5 & , \ \ & \tilde{q}_2 & = & p'_2 p'_7 \\
M_{11} & = & p'_1 p'_2 p'_3 & , \ \ & \tilde{q}_1 & = & p'_5 p'_7 
\end{array} \right\} \ \ F_{q_1}: \ \ (p'_1 p'_3 p'_5)(p'_2 p'_7) = (p'_1 p'_2 p'_3)(p'_5 p'_7)
\eeq
All other equations in \eref{F_mag} are similarly satisfied.

As for the electric theory, the information in the matrix \eref{P_magnetic} can be recast in terms of the characteristic polynomial

\beq
\begin{array}{ccl}
\mathcal{P}_m & = & M_{11}M_{12}M_{21}M_{22}+\mathcal{O}_1 \mathcal{O}_2 M_{21} M_{22} \tilde{q}_1 + \mathcal{O}_1 \mathcal{O}_4 M_{11} M_{21} q_2 + \mathcal{O}_2 \mathcal{O}_3 M_{12} M_{22} q_1 \\
& + & \mathcal{O}_3 \mathcal{O}_4 M_{11} M_{12} \tilde{q}_2 +  \mathcal{O}_1 \mathcal{O}_2 \mathcal{O}_3 \mathcal{O}_4 (q_1 q_2 + \tilde{q}_1 \tilde{q}_2).
\end{array}
\label{polynomial_magnetic}
\eeq

\bigskip

\subsection{Connecting Regular Parametrizations}

\label{section_connecting_parametrizations}

Let us now consider how the regular, perfect matching, parametrizations of the master spaces of the electric and magnetic theories are related. In order to do so, it is convenient to study the characteristic polynomials. The only objects that are not affected by the transformation and hence are present in both theories are the $\mathcal{O}_\alpha$ operators. Then, we have to look at \eref{polynomial_electric} and \eref{polynomial_magnetic} and compare terms with the same $\mathcal{O}_\alpha$ content. This is summarized in the \eref{polynomial_terms}, where we have included a relative normalization coefficient $\Lambda$.

\beq
\begin{array}{|c|c|c|}
\hline
& \mbox{Electric} & \mbox{Magnetic} \\ \hline \hline
1 & \ \ \ Q_1 Q_2 + \tilde{Q}_1 \tilde{Q}_2 \ \ \ & \ \ \ \Lambda \times M_{11} M_{12} M_{21} M_{22} \ \ \ \\ \hline
\mathcal{O}_1 \mathcal{O}_2 & Q_2 & \Lambda \times M_{21} M_{22} \, \tilde{q}_1 \\ \hline
\mathcal{O}_1 \mathcal{O}_4 & \tilde{Q}_1 &\Lambda \times M_{11} M_{21} \, q_2 \\ \hline
\mathcal{O}_2 \mathcal{O}_3 & \tilde{Q}_2 & \Lambda \times M_{12} M_{22} \, q_1 \\ \hline
\mathcal{O}_3 \mathcal{O}_4 & Q_2 & \Lambda \times M_{11} M_{12} \, \tilde{q}_2 \\ \hline
\ \ \mathcal{O}_1 \mathcal{O}_2 \mathcal{O}_3 \mathcal{O}_4 \ \ & 1 & \Lambda \times  (q_1 q_2 + \tilde{q}_1 \tilde{q}_2) \\ \hline
\end{array}
\label{polynomial_terms}
\eeq

\medskip

Equating the terms on the first row determines the normalization

\beq
\Lambda = {Q_1 Q_2 + \tilde{Q}_1 \tilde{Q}_2 \over M_{11} M_{12} M_{21} M_{22}}.
\label{cluster_normalization}
\eeq
The next four equations give

\beq
\begin{array}{cclcccl}
q_1 & = & {1 \over \Lambda \, M_{12} M_{22}} \, \tilde{Q}_2 & \ \ \ \ \ \ & 
\tilde{q}_1 & = & {1 \over \Lambda \, M_{21} M_{22}} \, Q_2 \\ \\
q_2 & = & {1 \over \Lambda \, M_{11} M_{21}} \, \tilde{Q}_1 & \ \ \ \ \ \ & 
\tilde{q}_2 & = & {1 \over \Lambda \, M_{11} M_{12}} \, Q_1 
\end{array}
\label{cluster_qi}
\eeq
\smallskip
\noindent Equations \eref{cluster_normalization} and \eref{cluster_qi} imply the terms in the last line of \eref{polynomial_terms} also match.


Equating the characteristic polynomials is a formal operation to connect the perfect matching parametrizations given by \eref{P_matrix_electric} and \eref{P_magnetic}. The physical meaning of relating terms with the same $\mathcal{O}_\alpha$ content is the following. It is straightforward to trade external faces in the graphs above for the (sets of) external legs separating consecutive pairs, i.e. for the $\mathcal{O}_\alpha$ operators. This means that the $\mathcal{O}_\alpha$ content of a term determines the transformation properties of the corresponding perfect matching under the global symmetries of the gauge theory, which are invariant under the duality. The simple example we are considering also exhibits a general phenomenon which also appears in more involved theories: the identification can occur between sets with a different number of perfect matchings. This is the case for the first line in \eref{polynomial_terms}, with maps the pair $p_6$, $p_7$ to $p_1'$.

The translation between the electric and magnetic sets of perfect matchings automatically ensures the matching of all gauge invariant operators in the chiral ring of the dual theories. Let us illustrate this in an explicit example, given by $Q_1 \tilde{Q}_1$ in the electric theory and $M_{11}$ in the magnetic one. We have:

\beq
\begin{array}{lccccc}
\mbox{{\bf Electric:}} & \ \ Q_1 \tilde{Q}_1 \ \ & = & (p_5p_6)(p_3 p_7) & = & [p_6p_7]_1 \, [p_3] _{\mathcal{O}_1\mathcal{O}_4} \, [p_5]_{\mathcal{O}_3\mathcal{O}_4}  \\
\mbox{{\bf Magnetic:}} & M_{11} & = & p'_1 p'_2 p'_3 & = & [p'_1]_1\, [p'_3] _{\mathcal{O}_1\mathcal{O}_4} \, [p'_2]_{\mathcal{O}_3\mathcal{O}_4}  
\end{array}
\eeq

\noindent where we have used \eref{P_matrix_electric} and \eref{P_magnetic}. In the last terms we have grouped perfect matchings according to the corresponding $\mathcal{O}_\alpha$ combinations, which are indicated by the subindices. In order to go from the electric to the magnetic theory, we simply need to replace the perfect matchings for each $\mathcal{O}_\alpha$ content for the corresponding ones in the dual.


The master space of a BFT can be parametrized in terms of paths on the graph \cite{Franco:2012wv}. For theories on a disk, it is sufficient to consider the set of loops around faces. In addition, in this case, face variables are directly related to coordinates in the Grassmannian.

Our goal now is to understand the effect of square moves on face variables. Before proceeding, let us discuss in more detail the problem at hand. We want to relate edge weights, which are identified with vacuum expectation values for scalar components of chiral fields, to oriented paths on graphs, of which face variables are particular examples. To do so, it is necessary to endow edge weights with an orientation since, strictly speaking, they do not have an inherent one (not to be confused with the bifundamental orientation). We will refer to the resulting objects as {\bf oriented edge weights}. In a slight abuse of notation, which is frequent in both the mathematics and physics literature, we will continue using the name of the corresponding edge to refer to the oriented edge weights. This should not lead to confusions since it is clear that whenever the discussion involves oriented paths, we refer to the latter. One possible systematic convention for assigning orientations is that oriented edge weights go from white to black nodes. With this prescription (see e.g. \cite{Kenyon:2003uj,Franco:2006gc} for applications of this idea), we can write any path on the bipartite graph in terms of the oriented edge weights as follows

\beq
v(\gamma)=\prod_{i=1}^{k-1} {X(\ww_i,\bb_i)\over X(\ww_{i+1},\bb_i)} \, ,
\label{flux_v}
\eeq
where the product runs over the path $\gamma$ and $\bb_i$ and $\ww_j$ denote black and white nodes. Here, $X(\ww_i,\bb_i)$ and $X(\ww_{i+1},\bb_i)$ are oriented edge weights, for which we explicitly indicate the graph nodes connected by the corresponding edge when moving along $\gamma$ instead of employing the usual notation with subindices for the two faces they separate. Going back and forth between the two notations is straightforward.

We are now in a position to explain in more detail the meaning of the $\mathcal{P}_\mu$'s in this context, they are {\bf oriented perfect matchings}. An oriented perfect matching is given by the product of the oriented weights for all edges in the corresponding perfect matching $p_\mu$, as determined by \eref{p_of_X}. These objects can be combined to form oriented paths.

The relations in \eref{cluster_qi}, which were obtained from formally equating terms in the characteristic polynomial, should not be understood as operator relations in the chiral ring. They are, however, relations between oriented weights, i.e. relation that can be used to determine the transformation of oriented paths in the graph under square moves. Using \eref{cluster_qi} and \eref{flux_v}, the face variables in the electric and magnetic theories are given by
\beq
\begin{array}{cclcccl}
W & = & {\tilde{Q}_1 \tilde{Q}_2 \over Q_1 Q_2} & \ \ \ \ \ \ & W' & = & {\tilde{q}_1 \tilde{q}_2 \over q_1 q_2} \\
W_1 & = & W_1^* Q_1 & & W_1' & = & W_1^* {M_{11} M_{12} \over \tilde{q}_1} \\
W_2 & = & W_2^* {1\over \tilde{Q}_1} & & W_2' & = & W_2^* {q_1 \over M_{21} M_{11}} \\
W_3 & = & W_3^* Q_2 & & W_3' & = & W_3^* {M_{22} M_{21} \over \tilde{q}_2} \\
W_4 & = & W_4^* {1\over \tilde{Q}_2} & & W_4' & = & W_4^* {q_2 \over M_{12} M_{22}}
\end{array}
\label{W_variables}
\eeq
Where $W_i^*$, $i=1,\ldots,4$, indicate the pieces of the face variables that do not include the edges involved in the square move. Plugging \eref{cluster_qi} into \eref{W_variables}, we obtain

\beq
\begin{array}{cclcccl}
W_1' & = & W_1 (1+W) & \ \ \ \ \ \ & W_2' & = & W_2 (1+W^{-1})^{-1} \\ 
W_3' & = & W_3 (1+W) & \ \ \ \ \ \ & W_4' & = & W_4 (1+W^{-1})^{-1} 
\end{array}
\eeq
These are precisely the cluster transformations of face weights \cite{Postnikov_plabic}. We have derived them exclusively at the level of the BFT gauge theory from matching regular parametrizations of the master spaces of the electric and magnetic theories. This behavior under square moves is in fact what is required for face weights to define coordinates of the totally non-negative Grassmannian.

A physical interpretation of face variables has been provided in \cite{ Heckman:2012jh}, where they have been related to vacuum expectation values of line operators in the 3d theories obtained by dimensional reduction of the BFT on a circle. A direct understanding of them in the 4d theory, beyond formal relations to other variables, is currently unavailable but would be highly desirable.

\bigskip

\section{Conclusions}

\label{section_conclusions}

In this paper we derived the cluster transformations of face weights exclusively in terms of BFTs. We showed that they follow from connecting regular parametrizations of the master space of theories related by square moves. This understanding complements the recent derivation of these transformations in terms of surface operators in dimensionally reduced BFTs \cite{Heckman:2012jh}. The results presented here are the latest addition to a list of structures associated to the combinatorics of the Grassmannian, which include the boundary operator on cells in the positive Grassmannian, and the matching and matroid polytopes, that have been identified in terms of BFT dynamics \cite{Franco:2012mm,Franco:2012wv}. 

\bigskip

\section*{Acknowledgments}

We would like to thank D. Galloni and R.-K. Seong for collaboration on related problems. We also thank N. Arkani-Hamed and J. Trnka for useful discussions. We are greatly indebted to N. Arkani-Hamed for posing the question leading to this paper and for comments on an earlier draft. This work is supported by the U.K. Science and Technology Facilities Council (STFC).

\bigskip

\appendix

\section{Comments on External Legs}

\label{appendix_legs}

In this appendix we collect a few remarks regarding external legs in BFTs. While these facts are rather clear in the literature, it might be useful to summarize them for future reference.  

\begin{figure}[h]
\begin{center}
\includegraphics[width=11cm]{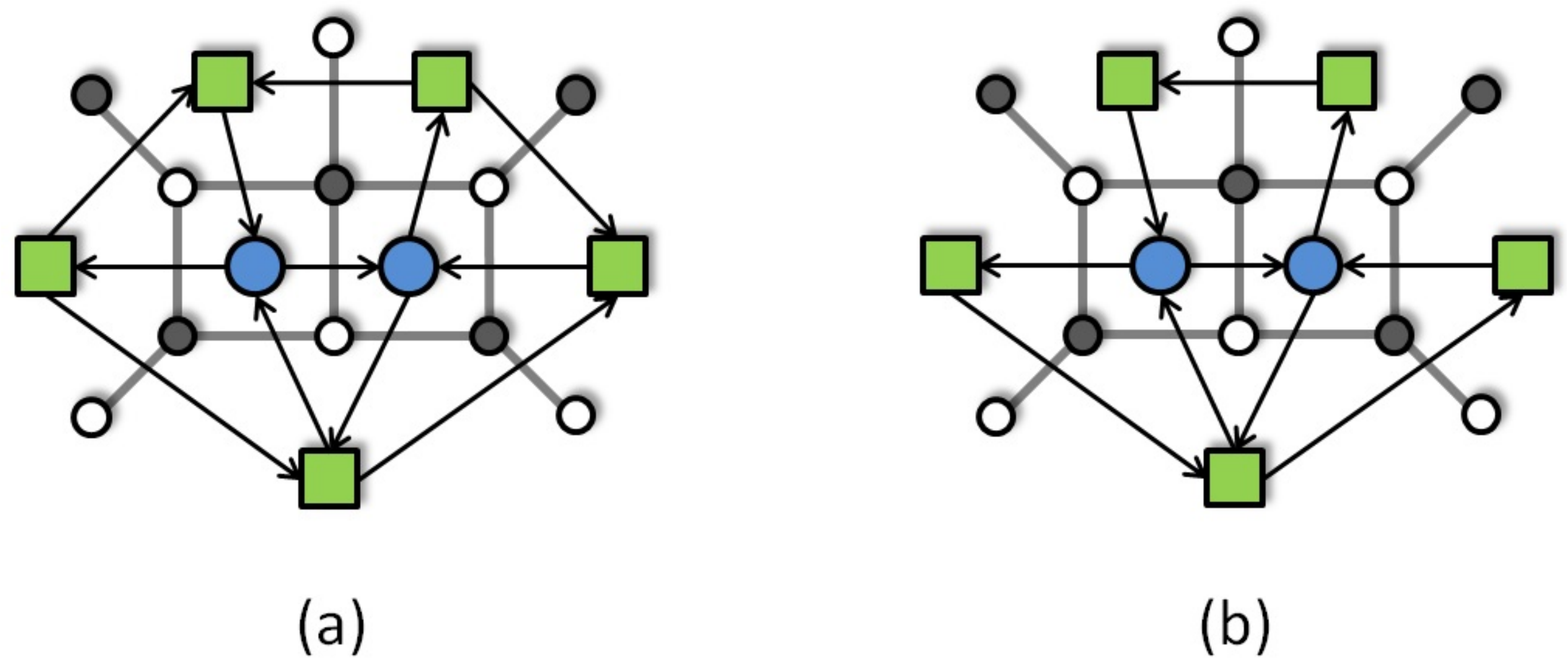}
\caption{\textit{Two prescriptions for assigning chiral fields to external legs.} We superimpose the quivers over the bipartite graph. (a) Every leg corresponds to a chiral field \cite{Franco:2012mm}. (b) Legs terminating on black external nodes are not associated to chiral fields. The corresponding plaquettes are missing from the superpotential \cite{Xie:2012mr}.}
\label{legs_BFT} 
\end{center}
\end{figure}

Let us first discuss the correspondence between edges in the bipartite graphs and chiral fields, for which two approaches have been proposed. The prescription in \cite{Franco:2012mm} associates a chiral field to every edge. On the other hand, according the rules in \cite{Xie:2012mr}, external legs connected to black external nodes do not have an associated chiral field. The superpotential terms associated to the missing plaquettes are also absent. \fref{legs_BFT} shows an example of the field content following each prescription. The connection between the two types of theories is straightforward. One can start from the theories in \cite{Franco:2012mm} and tune some of the superpotential couplings to zero, more precisely those associated to the white nodes connected to black external nodes. In this limit, the theories reduce to the ones in \cite{Xie:2012mr} plus decoupled singlets, which correspond to the legs connected to black external nodes.

Another issue, which is intimately related to the previous paragraph, is whether the fields for external legs are dynamical or not. Both possibilities have natural motivations and specific implications. In \sref{section_BFTs_nutshell}, several reasons for considering non-dynamical external legs were given. They range from the interest in making contact with the geometry of the Grassmannian to general considerations regarding the implementation of global symmetry groups in terms of higher dimensional flavor D-branes. We also explained that, with this prescription, square moves consistently implement Seiberg duality for any internal square face, i.e. including those adjacent to external faces \cite{Franco:2012mm}. 

Let us instead consider what happens when dynamical fields are assigned to external legs. In this case, the analysis of Seiberg duality on square faces adjacent to external ones is slightly different. The rule stating that legs connected to external black nodes are not associated to chiral fields becomes necessary for consistently incorporating such dualities \cite{Xie:2012mr}. This follows from the generation of mass terms along external legs. Summarizing the previous two paragraphs, the choice of map between chiral fields and external legs can thus be regarded as a result of whether they are dynamical or not. 

Let us now consider general properties of string theory implementations, in those cases in which they exist, of dynamical fields for external legs. As already mentioned, external faces are naturally associated to higher dimensional flavor branes. The fields under consideration would live at their intersections, which are also higher dimensional. In order for them to become dynamical in 4d, they must have finite support in the internal dimensions.  This can be simply achieved by cutting off the intersections between flavor branes. Quite generally this requires cutting off the flavor branes themselves, resulting in the gauging of the corresponding symmetries. The gauging of external faces might be arbitrarily weak when compared to the one of internal faces. We refer the reader to \cite{Heckman:2012jh} for an explicit realization of this scenario. The internal volume of flavor branes can indeed grow very rapidly with the cut off, but will remain finite for any finite value of it. Nevertheless, this gauging imposes an important constraint on the BFT: external faces must be anomaly free. It is straightforward to see that the assignation of fields to external legs of \fref{legs_BFT} (b) provides a general way of satisfying this condition for any bipartite graph. An alternative way of thinking about this assignment is that all external legs correspond to bifundamental fields, but that extra matter needs to be incorporated for anomaly cancellation. These additional fields have the right quantum numbers to couple via mass terms to the legs that end on black external nodes, lifting them at low energies.

It is amusing that one can show that the assumption of all fields in the BFT being dynamical leads to \fref{legs_BFT} (b) using two seemingly independent arguments. One of them is based on Seiberg duality and the other one follows from general properties expected in string theory embeddings of these theories. As we explained, there are also reasons for not considering these fields as dynamical.

We hope our discussion clarifies the relation between the treatments in \cite{Franco:2012mm} and \cite{Xie:2012mr}, their motivations and how simple it is to switch between them.

\bigskip
\bigskip



\begin{thebibliography}{10}

\bibitem{Franco:2012mm} 
  S.~Franco,
  ``Bipartite Field Theories: from D-Brane Probes to Scattering Amplitudes,''  arXiv:1207.0807 [hep-th].  

\bibitem{Xie:2012mr} 
  D.~Xie and M.~Yamazaki,
  ``Network and Seiberg Duality,''  JHEP {\bf 1209}, 036 (2012)  [arXiv:1207.0811 [hep-th]].  

\bibitem{Hanany:2005ve} 
  A.~Hanany and K.~D.~Kennaway,
  ``Dimer models and toric diagrams,''  hep-th/0503149.  

\bibitem{Franco:2005rj} 
  S.~Franco, A.~Hanany, K.~D.~Kennaway, D.~Vegh and B.~Wecht,
  ``Brane dimers and quiver gauge theories,''  JHEP {\bf 0601}, 096 (2006)  [hep-th/0504110].  

\bibitem{Franco:2005sm} 
  S.~Franco, A.~Hanany, D.~Martelli, J.~Sparks, D.~Vegh and B.~Wecht,
  ``Gauge theories from toric geometry and brane tilings,''  JHEP {\bf 0601}, 128 (2006)  [hep-th/0505211].  

\bibitem{Butti:2005sw} 
  A.~Butti, D.~Forcella and A.~Zaffaroni,
  ``The Dual superconformal theory for L**pqr manifolds,''  JHEP {\bf 0509}, 018 (2005)  [hep-th/0505220].  

\bibitem{Feng:2005gw} 
  B.~Feng, Y.~-H.~He, K.~D.~Kennaway and C.~Vafa,
  ``Dimer models from mirror symmetry and quivering amoebae,''  Adv.\ Theor.\ Math.\ Phys.\  {\bf 12}, 489 (2008)  [hep-th/0511287].  

\bibitem{GK}
A.~Goncharov and R.~Kenyon, 
``Dimers and cluster integrable systems,''
arXiv:1107.5588 [math.AG]

\bibitem{Franco:2011sz} 
  S.~Franco,
  ``Dimer Models, Integrable Systems and Quantum Teichmuller Space,''  JHEP {\bf 1109}, 057 (2011)  [arXiv:1105.1777 [hep-th]].  

\bibitem{Eager:2011dp} 
  R.~Eager, S.~Franco and K.~Schaeffer,
  ``Dimer Models and Integrable Systems,''  arXiv:1107.1244 [hep-th].  

\bibitem{Franco:2012hv} 
  S.~Franco, D.~Galloni and Y.~-H.~He,
  ``Towards the Continuous Limit of Cluster Integrable Systems,''  JHEP {\bf 1209}, 020 (2012)  [arXiv:1203.6067 [hep-th]].  

\bibitem{Nima} 
  N.~Arkani-Hamed, J.~L.~Bourjaily, F.~Cachazo, A.~B.~Goncharov, A.~Postnikov and J.~Trnka,
  ``Scattering Amplitudes and the Positive Grassmannian,''  arXiv:1212.5605 [hep-th].  

\bibitem{ArkaniHamed:2009dn} 
  N.~Arkani-Hamed, F.~Cachazo, C.~Cheung and J.~Kaplan,
  ``A Duality For The S Matrix,''  JHEP {\bf 1003}, 020 (2010)  [arXiv:0907.5418 [hep-th]].  

\bibitem{Postnikov_plabic} 
  A.~Postnikov,
  ``Total positivity, Grassmannians, and networks,''  [arXiv:math/0609764 [math.CO]].

\bibitem{Postnikov_toric} 
  A.~Postnikov, D.~Speyer, L.~Williams,
  ``Matching polytopes, toric geometry, and the non-negative part of the Grassmannian,''  [arXiv:0706.2501 [math.AG]].

\bibitem{Franco:2012wv} 
  S.~Franco, D.~Galloni and R.~-K.~Seong,
  ``New Directions in Bipartite Field Theories,''  arXiv:1211.5139 [hep-th].  

\bibitem{Heckman:2012jh} 
  J.~J.~Heckman, C.~Vafa, D.~Xie and M.~Yamazaki,
  ``String Theory Origin of Bipartite SCFTs,''  arXiv:1211.4587 [hep-th].  

\bibitem{FZ4}
S.~Fomin and A.~Zelevinsky, ``Cluster algebras IV: Coefficients," Compos. Math. 143 (2007), 112--164. 

\bibitem{Propp}
J.~Propp, ``Generalized domino-shuffling,'' Theoret. Comput. Sci. 303, no. 2-3, 267--301, Tilings of the plane (2003) [arXiv:math/0111034 [math.CO]].

\bibitem{Forcella:2008bb} 
  D.~Forcella, A.~Hanany, Y.~-H.~He and A.~Zaffaroni,
  ``The Master Space of N=1 Gauge Theories,''  JHEP {\bf 0808}, 012 (2008)  [arXiv:0801.1585 [hep-th]].  

\bibitem{Kenyon:2003uj}
  R.~Kenyon, A.~Okounkov and S.~Sheffield,
  ``Dimers and Amoebae,''
  arXiv:math-ph/0311005.

\bibitem{Franco:2006gc} 
  S.~Franco and D.~Vegh,
  ``Moduli spaces of gauge theories from dimer models: Proof of the correspondence,''  JHEP {\bf 0611}, 054 (2006)  [hep-th/0601063].  

\bibitem{Seiberg:1994pq} 
  N.~Seiberg,
  ``Electric - magnetic duality in supersymmetric nonAbelian gauge theories,''  Nucl.\ Phys.\ B {\bf 435}, 129 (1995)  [hep-th/9411149].  

\bibitem{MR1887642}
S.~Fomin and A.~Zelevinsky, ``Cluster algebras I. Foundations,"  {\em
  J. Amer. Math. Soc.} {\bf 15} (2002), no.~2 497--529 (electronic).


\end{thebibliography}


\end{document}